\documentstyle [12pt] {article}

\catcode`@=12
\newcommand{\be}{\begin{equation}}
\newcommand{\ee}{\end{equation}}
\newcommand{\ber}{\begin{eqnarray}}
\newcommand{\eer}{\end{eqnarray}}
\newcommand{\bers}{\begin{eqnarray*}}
\newcommand{\eers}{\end{eqnarray*}}
\newcommand{\bt}{\begin{itemize}}
\newcommand{\et}{\end{itemize}}
\begin{document}
\vspace{0.5in}
\oddsidemargin -.375in  
\newcount\sectionnumber 
\sectionnumber=0 
\def\be{\begin{equation}} 
\def\ee{\end{equation}}
\thispagestyle{empty}  
\begin{flushright} UTPT-98-15 \\September 1998\
\end{flushright}
\vspace {.5in} 
\begin{center} 
{\Large\bf Implications of Isospin Conservation in $\Lambda_b$\\  
Decays and Lifetime \\}
\vspace{.5in}
{{\bf Alakabha Datta{\footnote{email: datta@medb.physics.utoronto.ca}}
${}^{a}$}, {\bf Harry J. Lipkin {\footnote{email: hjl@hep.anl.gov}}
 ${}^{b}$} and
{\bf Patrick. J. O'Donnell {\footnote{email: pat@medb.physics.utoronto.ca}}${}^{a}$} \\}
\vspace{.1in} 
${}^{a}$ {\it Department of Physics and Astronomy,\\  
University of Toronto, Toronto, Canada.}\\  
${}^{b)}$ {\it 
Department of Particle Physics,\\
Weizmann Institute,\\
Rehovot 76100, Israel \\and\\
School of Physics and Astronomy, \\
Tel-Aviv University,\\ 
Tel-Aviv 69978, Israel \\  
}
\end{center}

\begin{abstract}
 We consider isospin predictions for the semi-leptonic  and non-leptonic decays
of the $\Lambda_b$ baryon. Isospin conservation of the strong interactions constrains the 
possible final states in $\Lambda_b$ decays. This leads in general to phase space 
enhancements in $\Lambda_b$ decays  relative to $B$ meson decays for the same underlying quark transitions. Consequently the $\Lambda_b$ lifetime is smaller than the $B$ lifetime. Phase space enhancements in $\Lambda_b$ decays 
relative to $B$ decays can be understood in
terms of hyperfine
interactions in the bottom system.
\end{abstract}

\newpage \pagestyle{plain}

\section{Introduction}
Isospin conservation is a good approximate symmetry of the strong interactions
and can be applied fruitfully in $\Lambda_b$ decays. The $\Lambda_b$ baryon
is made of a heavy $b$ quark and a $ud$ light diquark system 
in a spin and isospin
singlet state. Semi-leptonic $\Lambda_b$ decays involve the weak
$b \to c$ transition without the involvement of the light quarks.
The final hadronic decay products have to be in an isosinglet state
as the weak current is an isoscalar. 
Strong interactions do not change the isospin state of the light diquark
which combines with the $c$ quark to form the hadrons in the final state.
Single particle hadronic states would therefore dominantly involve the ground and excited
$\Lambda_c$ baryons.
In non-leptonic $\Lambda_b$ decays the
 effective current$\times$current Hamiltonian gives rise
to the following quark diagrams  \cite{CH} : the internal and
external W-emission diagrams, which
result in the factorizable contribution, and the W-exchange diagrams which
gives
rise to the non-factorizable contribution.
The W-annihilation diagram is absent in
baryon decay and we neglect the penguin contributions. The 
contribution from the 
W-exchange diagram 
is expected to be small in $\Lambda_b$ decays. The final states 
in non-leptonic
$\Lambda_b$ decays result from the isosinglet diquark combining with the
final state quarks. For instance
 in the quark level transition $ b \to c {\overline c }s$ the diquark 
can combine with the $c$ quark in the final state. 
Hence final states like ${\overline D_s} \Lambda_c$ are allowed but 
states like ${\overline D_s} \Sigma_c$ are not. 
 
As the quark mass  becomes  heavier  many  differences  among the
properties of spin--$1/2$ and spin--$3/2$  baryons and also among
pseudo scalar   and  vector  mesons  containing  a heavy  quark are
expected  to  become  less  pronounced   \cite{Neubert}. 
  As the quark mass  increases  it is expected  that
the lifetimes of particles containing one heavy quark will become
very similar \cite{Bigi}.  It is in the corrections to the lowest
order in $\Lambda_{QCD}/M$ where models play a role.

In Heavy Quark Effective Theory (HQET) the lifetimes of the $\Lambda_{b}$ and the
$B^0$  meson  were  expected  to be the same in the heavy
quark  limit  and just  slightly  different  when  certain  quark
scattering  processes that could occur in the  $\Lambda_{b}$  but
not in the meson were included.  These  principally  included (a)
the  ``weak   scattering"   process,   first   invoked   for  the
$\Lambda_{c}^+$  lifetime  \cite{Barger},  and here of the  form,
$bu\rightarrow cd$, and (b), the so--called ``Pauli interference"
process  $bd\rightarrow  c\bar u dd$  \cite{Voloshin,Bilic}.  The
results of including  these terms is a slight  enhancement in the
decay rate  leading  to  $\tau(\Lambda_{b})/\tau(B^0)  \sim 0.9$,
whereas the evaluation \cite{pdg} of $\tau(\Lambda_{b})$ is $1.24
\pm 0.08$ ps and $\tau(B^0) = 1.56 \pm 0.04$ ps gives a very much
reduced fraction  $\tau(\Lambda_{b})/\tau(B^0)  = 0.79 \pm 0.07$,
or  conversely  a very much  enhanced  decay  rate.  (There  is a
recent CDF  result  \cite{CDF}  which  would  move this  fraction
higher  than the world  average  to a value of $0.85 \pm 0.10 \pm
0.05$).

A possible explanation of the $\Lambda_b$ lifetime is an
 enhancement of the decay width, $\Delta  \Gamma(\Lambda_{b})$
from the $q - q$  scattering. This  involves  replacing  the usual flux
factor by  $|\psi(0)|^2$,  the wave function at the origin of the
pair of quarks $bu$ in the  $\Lambda_{b}$, (or the pair $bd$, for
which the wave  function is the same by isospin  symmetry).  This
wave  function  at the  origin  naturally  appears  in  hyperfine
splitting  \cite{dgg}.  Rosner \cite{Rosner} tried to account for
the enhancement by changing the wave function $|\psi(0)_{bu}|^2$; this would 
also correlate with the surprisingly large hyperfine splitting suggested by the 
DELPHI group \cite{DELPHI}.
He was able to show that, under certain  assumptions, there could
be at most a  $13\pm  7 $\%  increase  of the  amount  needed  to
explain the decay rate of the $\Lambda_{b}$.
In a more dramatic attempt to explain the lifetime problem it has
been    proposed    \cite{NS}   to   allow   the   ratio   $r   =
|\psi_{bq}^{\Lambda_b}(0)|^2/   |\psi_{b\bar  q}^{B_q}(0)|^2$  to
vary between $1/4$ and $4$.  Clearly such a large variation would
be ruled out by hyperfine relations.

Here we show that isospin conservation leads naturally to a phase space enhancement
in $\Lambda_b$ decays relative to $B$ decays resulting in a shorter lifetime
for $\Lambda_b$. As shown in \cite{LO} isospin conservation chooses the final state
in $\Lambda_b$ decays with the lowest hyperfine energy. In $B$ decays the spectator quark can combine with the $c$ quark to form vector or pseudo scalar final states. The hyperfine
energy in this case is averaged out resulting in a phase space advantage for  the baryon transition over the meson transition.

 In the following sections we study the isospin predictions in semi-leptonic and non-leptonic
$\Lambda_b$ decays. We then show that phase space enhancements in $\Lambda_b$
decays relative to $B$ decays lead to shorter lifetime for $\Lambda_b$ relative to $B$.

\section{Semi-Leptonic Decays}
 Semi-leptonic $\Lambda_b$ decay involves the quark level $b \to c$ transition
due to an isoscalar current.
The amplitude for the process can be written as
\ber
A & = & <X| J_{\mu} |\Lambda_b>L^{\mu} \
\eer
where $J_{\mu} = {\overline c}\gamma_{\mu}(1-\gamma_5)b$ is the isoscalar weak current and $L^{\mu}$ is the leptonic weak current. The final state $X$ 
has to be in an isosinglet state.
In the heavy quark limit the light degrees of freedom in a hadron, the 
diquark in this case, have conserved isospin
and angular momentum quantum numbers.
Due to isospin conservation the light diquark in $\Lambda_b$ 
remains in an isosinglet
state as it combines with the $c$ quark to generate the spectrum 
of final states. When the diquark combines with the $c$ quark it will form
dominantly a $\Lambda$ type charmed baryon. The $\Lambda$ type baryon 
can be classified according to the quantum numbers carried  
by the light degrees of freedom. So the lowest state corresponds 
to the light degree having isospin $I_l=0$ and spin $s_l=0$. The diquark 
can be excited to a higher orbital angular momentum state with $L_l=1$. This 
creates baryons with net spin $1/2$ and $3/2$ denoted by 
$\Lambda^*_{c1}$
and $\Lambda^*_{c2}$ or alternately as $\Lambda_c(2593)$ and $\Lambda_c(2625)$.
 Other final states that can populate the $X$ spectrum to a lesser fraction
 than single particle $\Lambda_c$ states
are $D^0 p$ $D^0 p \pi^0$ etc. Note isosinglet combinations like 
$\Sigma_c^{++}\pi^-$, $\Sigma_c^{+}\pi^0$ can only be the decay product of
excited $\Lambda_c$ type baryons where the pion is emitted from the diquark
changing it from an isosinglet to an isovector state.

Hence the decay $\Lambda_b \to X l {\overline \nu}$ should be dominantly
$\Lambda_b \to \Lambda^{(*)} l {\overline \nu}$ where $\Lambda^{(*)}$ 
denotes the ground
state or the excited $\Lambda_c$. Because the 
excited $\Lambda_c$ decays to the ground state
$\Lambda_c$ we have the prediction
$$ \Lambda_b \to X l {\overline \nu} \approx \Lambda_b \to \Lambda_c X l {\overline \nu}.$$
  We can therefore have the following decays for $\Lambda_b \to X l {\overline \nu}$ :
\bt
\item{
$\Lambda_b \to \Lambda_c l {\overline \nu}$. This is
 expected to be the dominant decay in the inclusive 
semi-leptonic $\Lambda_b$ decay. In the heavy quark limit at 
maximum $q^2$ where $q^2$ is the invariant energy of the 
$l {\overline \nu}$ system or
equivalently at $\omega=v \cdot v'=1$ where $v$ and $v'$ are the initial 
and final baryon velocities transitions to excited $\Lambda_c$ are suppressed by $1/m_b^2$.
Model estimates of this branching fraction are between $7 - 8 \%$ \cite{Gupta} while 
estimate of $\Lambda_b \to \Lambda_c X l {\overline \nu}$ is around $(10 \pm 4)\%$
\cite{pdg}. This also indicates that $\Lambda_c \to \Lambda_c l {\overline \nu}$ dominates
$\Lambda_b \to X l {\overline \nu}$ .
}   

\item {
\bers
\Lambda_b & \to & \Lambda_c(2593,2625) l {\overline \nu}\\
\Lambda_c(2593,2625) & \to & \Lambda_c \pi^+ \pi^- \\
          & \to & \Lambda_c \pi^0 \pi^0 \\
          & \to & \Sigma_c(2455)^{++} \pi^- {\stackrel{100\%}{\to}}\Lambda_c \pi^+ \pi^- \\
          & \to & \Sigma_c(2455)^{0} \pi^+ {\stackrel{100\%}{\to}}\Lambda_c \pi^- \pi^+ \\
\eers
From the above we see that if a $\Sigma_c$ is in the final state 
it must be associated with a $\pi$ and further the invariant mass of 
the $\Sigma_c - \pi$ or the $\Lambda_c \pi \pi$
 system must be the mass of the excited $\Lambda_c$.
}
\et
Finally our prediction is
$$ \Lambda_b \to X l {\overline \nu} \approx 
\Lambda_b \to \Lambda_c X l {\overline \nu}
=\Lambda_b \to \Lambda_c l {\overline \nu},  
 \Sigma_c \pi l {\overline \nu},  
\Lambda_c \pi \pi l {\overline \nu}$$

In semi-leptonic $B$ decays the largest fraction
of final
states 
will involve the $D$ and the $D^*$ meson. When compared to the dominant decay
$\Lambda_b \to \Lambda_c l {\overline \nu}$ there is a phase space advantage
in $\Lambda_b$ decays relative to $B$ decays which results in a shorter 
lifetime for $\Lambda_b$ relative to $B$. We will discuss this more 
quantitatively in a later section.
\section{Non-Leptonic Decays}
Non-leptonic $\Lambda_b$ decays proceed through the underlying quark 
transitions $b \to c {\overline c} s'$ 
and $b \to c {\overline u} d'$
where $d'=d \cos \theta_c
 + s \sin \theta_c$ and
$s'=-d \sin \theta_c + s \cos \theta_c $. We neglect the $b \to u$
and penguin transitions. As mentioned in the introduction
non-leptonic transitions involve the W-emission and the W 
exchange diagrams. From the study of 
$\Lambda_b$ lifetime, the W-exchange contribution 
relative to the spectator $b$ quark decay rate is of the order 
$32 \pi^2 |\psi(0)|^2/m_b^3$. This is of the order unity in the case of
 charmed baryons \cite{Bilic,Lcheng} ( which has $m_c$ in place of $m_b$ ) and so 
is much suppressed in the case of $\Lambda_b$ baryons. Note the wave 
function at the origin, $\psi(0)$, is approximately same for the 
charm and bottom system. Hence in non-leptonic 
$\Lambda_b$ decays the W-exchange term will be small unlike in the case of 
charmed baryons. So factorization is expected to be a 
good approximation in the study of non-leptonic $\Lambda_b$ decays.

We now list the predictions for $\Lambda_b$ non-leptonic decays which 
follow from the  conservation of the isospin quantum number of the 
light diquark in the $\Lambda_b$ baryon
\bt
\item{
For $b \to c {\overline c} s$ transition the effective Hamiltonian is
\ber
H_W & = & c_1 {\overline c}b{\overline s}c + 
c_2{\overline s}b {\overline c}c \
\eer
where $c_{1,2}$ are the Wilson's coefficients and we have suppressed the 
color and Dirac index as well as the $\gamma_{\mu}(1-\gamma_5)$ factors. 
Since the W-emission diagram, which is given by the factorization amplitude,
is the dominant contribution here
we can write the non-leptonic amplitude as
\ber
A[\Lambda_b \to X X^{\prime}] & = &(c_1 +c_2/N_c)
<X|{\overline c}b|\Lambda_b><X^{\prime}|
{\overline s } c|0> \\
A_s[\Lambda_b \to X X^{\prime}] & = &(c_2 +c_1/N_c)
<X|{\overline s}b|\Lambda_b><X^{\prime}|
{\overline c } c|0> \
\eer
where $A$ and $A_s$ are the color allowed and color 
suppressed amplitudes and $N_c$ is the number of colors.
Now for the color allowed transition, from
our analysis of the semi-leptonic decays, $X$ is mainly 
$\Lambda_c^{(*)}$. Hence some possible final states are
$\Lambda_c {\overline D_s}$, 
$\Sigma_c \pi {\overline D_s}$, 
$\Lambda_c \pi \pi {\overline D_s}$. 
Note no 
single $\Sigma_c (\Sigma_c^*)$  is possible in the final state 
unless accompanied
by a pion. For the color suppressed transitions we can have final states like
$\Lambda^{(*)} J/\psi(D {\overline D})$, $\Sigma(\Sigma^*)
 \pi J/\psi(D {\overline D})$. 
Again  a single $\Sigma (\Sigma^*)$ final state is not allowed unless 
accompanied by a pion. Also most of the time  the
 $\Sigma \pi$ system would be the 
decay product of an excited $\Lambda$ and therefore would have an 
invariant mass of an excited $\Lambda$.

In $B$
 decays the same Hamiltonian would generate $D(D^*) {\overline D_s}$ final 
states in color allowed transitions and and 
as in the semi-leptonic case when we sum over all $|X^{\prime}>$ states
there will be an enhancement in the $\Lambda_b$ width relative to the $B$ 
width.
}
\et
\bt
\item{
The Cabibbo suppressed $ b\to c d {\overline d}$ color allowed 
transitions would give rise to the following possible final states
$\Lambda_c {\overline D^{(*)}}$, 
$\Sigma_c \pi {\overline D^{(*)}}$, 
$\Lambda_c \pi \pi {\overline D^{(*)}}$. Color suppressed transitions 
would have states like 
$N^{(*)} J/\psi(D {\overline D})$, 
$\Delta \pi J/\psi(D {\overline D})$. A single $\Delta$ in the final 
state is disallowed unless accompanied by a pion and in 
most cases the $\Delta-\pi$ invariant mass would  correspond to an excited 
nucleon.}
\et
\bt
\item{
Cabibbo allowed $b \to c {\overline u}d$ can lead to final states as
$\Lambda_c^{(*)} \pi(\rho)$, 
$\Sigma_c \pi \pi(\rho)$, 
$\Lambda_c \pi \pi \pi(\rho)$. Color suppressed decays will have final 
states as $ N^{(*)} D^0(D^{0*})$, 
$ \Delta \pi D^0(D^{0*})$.
} 
\et
\bt
\item{
Cabibbo suppressed $b \to c {\overline u}s$ can lead to final states as
$\Lambda_c^{(*)} K(K^*)$, 
$\Sigma_c \pi K(K^*)$, 
$\Lambda_c \pi \pi K(K^*)$. Color suppressed decays will have final 
states as $ \Lambda^{(*)} D^0(D^{0*})$, 
$ \Delta \pi D^0(D^{0*})$.
} 
\et
            
\section{$\Lambda_b$ Lifetime}
 Lifetimes of the $\Lambda_b$ and $B$ are calculated using the 
operator product expansion (O.P.E) to write the square of the decay amplitude
as a series of local operators \cite{NS}. The expression for the 
lifetime can be arranged
as an expansion in $1/m_b$. The inclusive rate calculated in this manner is
expected to equal the inclusive rate by summing up individual exclusive modes
by assumption of duality. The validity of duality has not been proved but 
it can be shown in a certain kinematic limit, the Shifman-Voloshin 
limit, defined by $m_b,m_c>>(m_b-m_c)>>\Lambda_{QCD}$ that the inclusive rate
calculated by the method of OPE gives the same result as summing up the exclusive modes which are saturated by $B \to D +D^*$ in $B$ decays and $\Lambda_b \to \Lambda_c$ in
$\Lambda_b$ decays \cite{SV,Manohar}.

 As mentioned in the introduction, in the 
leading order, the lifetimes of $\Lambda_b$ and $B$ are expected to be same if the OPE
method is used in calculating the lifetimes.
Spectator effects that distinguish between $\Lambda_b$ and $B$ only arise at
order $1/m_b^3$ and are not enough to explain the observed $\Lambda_b$, $B$
lifetime ratio.
In our analysis of exclusive  semi-leptonic decays we found that $\Lambda_b$ goes 
dominantly to $\Lambda_c$ while $B$ goes to $D$ and $D^*$. The result 
is a phase space advantage in the baryon transition over the meson transition
leading to an enhanced $\Lambda_b$ lifetime relative to the 
$B$ lifetime. We can calculate
the inclusive rates by summing up the exclusive modes.
For a  quantitative  estimate we use the following  toy model for
semileptonic decays:  We assume that the $\Lambda_b$ goes only to
$\Lambda_c$, that the $B$ goes to a statistical  mixture (3/4) D*
and  (1/4) D and  that  all  transitions  to  higher  states  are
small. In the SV limit, in the leading order, for semi-leptonic transition
$H_1 \to H_2 l {\overline \nu}$ the decay rates go as $(H_1-H_2)^5$. In
our toy model we will assume that this behavior of the decay  rate persists 
away from the SV limit also. Therefore
  the phase  space for the  $\Lambda_b$  decay is then
given by the mass difference $\Lambda_b - \Lambda_c$ to the fifth
power.  The  phase  space for the $B$ decay is then  given by the
$B-D^*$  mass  difference  to  the  fifth  power,  weighted  by a
statistical factor of (3/4) plus the $B-D$ mass difference to the
fifth power, weighted by a statistical factor of (1/4). It is interesting to note that
in this toy model $\Gamma(B \to D l {\overline \nu})/\Gamma(B \to D^* l {\overline \nu}) =0.41$
which is very close to the experimental number 0.42 \cite{pdg}.

Including small corrections from neglected transitions we can write
\ber
\Gamma_{SL}(\Lambda_b)& = & A(\Lambda_b -\Lambda_c)^5(1+x_1)\\
\Gamma_{SL}(B) &= &A(\frac{1}{4}(B-D)^5 +\frac{3}{4}(B-D^*)^5)(1+x_2)\
\eer
where $A$ is a constant involving the Fermi constant $G_F$ and $x_{1,2}$ 
are small corrections
from neglected transitions.
This well-defined  model for semi-leptonic  decays may be right or
wrong, but its  predictions  are easily  calculated and the basic
assumptions can be easily tested when exclusive  branching ratios
into baryon final states  including  spin-excited  baryons become
available.  We  immediately  obtain the following  result for the
ratio of semi-leptonic partial widths for $x_1 \approx x_2$:
\begin{equation}                                                                
{{\Gamma (\Lambda_b)}\over {\Gamma(B)}} =1.07 
\label{spectwidth} 
\end{equation}  
In a toy model including only semi-leptonic modes  this would
give the ratio of the lifetimes
\begin{equation}                                                                
{{\tau(\Lambda_b)}\over {\tau(B)}} =0.934        
\label{specttime} 
\end{equation}                          
This shows a clear prediction of a significant enhancement of the
$\Lambda_b$  partial  semi-leptonic  width in comparison  with the
$B$.  The $\Lambda_b$ decay rate is enhanced by about 7\%.

In the HQET picture of the 
hadrons, the heavy quark inside the hadron interacts with a complicated
``brown muck''. In the case of the meson there is only a single 
``brown muck'' in a isospin $I=1/2$ state while in the baryon the 
``brown muck'' can be in a isospin $I=0$ or $I=1$ state corresponding to the
$\Lambda$ and the $\Sigma$ baryon. In contrast to the meson hyperfine 
mass splittings, which are always between states of the same isospin and 
decrease to zero with the heavy quark mass, the $\Lambda-\Sigma$ 
splittings are between states of different isospins and therefore 
separated from one another by isospin selection rules.. Furthermore, they 
do not decrease with heavy quark mass, but actually increase, and are expected in simple models to approach a finite asymptotic value of $200$ MeV with infinite heavy quark mass \cite{Rujula}.
In the standard HQET expansion this spin-isospin splitting is neglected and 
as we have shown above, the effect of the spin-isospin 
splitting on phase space can be appreciable. 

One can also find evidence of a similar enhancement in 
non-leptonic decays. Consider for instance
the quark transition $b \to c {\overline u}d$. Considering only color 
allowed transitions
we found that for $\Lambda_b$ decays the final states are of the 
form $\Lambda_c^{(*)} X$
where $X=\pi \rho a_1 n\pi ...$. In the case of $B$ decays the final states 
are dominantly $D(D^*) X$.
If we now sum over the states $X$ then in the leading order 
we have the effective transitions
\bers
\Lambda_b \to \Lambda_c^{(*)} {\overline u}d \\
B \to D(D^*) {\overline u}d \
\eers

Here we have used the idea of duality 
in summing over the $X$ states. This 
maybe 
reasonable because there are
 many hadronic  channels and so summing over all the final states
will eliminate the bound state effects of the individual final states.
We can then apply the toy model for semi-leptonic decays considered above
 and we see that there is a phase space enhancements for $\Lambda_b$ 
decays relative to $B$ decays. A similar treatment can also be applied
 for other color allowed 
non-leptonic transitions taking proper care of the phase space factors. 
For instance in $\Lambda_b \to \Lambda_c {\overline u}d$ the invariant mass
$M_X$ varies from $(\Lambda_b -\Lambda_c)$ to $(m_u +m_d)$ and for
$\Lambda_b \to \Lambda_c {\overline c}s$ the invariant mass
$M_X$ varies from $(\Lambda_b -\Lambda_c)$ to $(m_c +m_s)$.

Note that in the traditional approach to calculating lifetimes using 
duality the isospin selection rules are not taken into account. For instance
 the transitions 
due to 
$b \to c {\overline c} s$ 
are $\Delta I =0$ 
and so the final 
states   in $\Lambda_b$ decays are 
rigorously required to be in an isoscalar state while in $B$ decays only
$I=1/2$ states are allowed. Using duality, in the leading order, both the 
$\Lambda_b$ and $B$ decays would be represented by the parton level process
$b \to c {\overline c} s$. The dynamics of the two light quarks in the baryon
and consequently the fine details of the hadron spectrum is ignored. While 
it is conceivable that the arguments supporting quark-hadron duality 
which neglect the fine details of the hadron spectrum maybe valid for 
mesons it is likely to break down for baryons where there are 
two valence light 
quarks undergoing very complicated non-perturbative QCD interactions.
We note that the use of duality even in the case of the $B$ meson has been questioned
recently \cite{Isgur}.

As a concrete example consider the color 
suppressed $b \to c {\overline c} s$  transition leading to
the processes $B \to J/\psi X$ and $\Lambda_b \to J/\psi X$. 
 Possible final 
states, as already mentioned before, for the $\Lambda_b$ decay  
are $J/\psi \Lambda^{(*)}$. In 
$B$ decays the corresponding final states are in $I=1/2$ states and some
 possible final states are $J/\psi K^{(*)}, K \pi \pi$ \cite{pdg}. If 
we used duality to sum over the $X$ states then in the leading order  both
$\Lambda_b$ and $B$ decaying to final state $J/\psi X$
 could be represented by $b \to s J/\psi$ and so the rates 
for both processes would be same. On the other hand isospin selects 
specific $X$ states. From measured rates in Particle Data Group \cite{pdg} 
$X=\Lambda$ and $K, K^{*}, K \pi \pi $ for 
$\Lambda_b$ and $B$ decays. If 
we add the observed rates we find $\Gamma[B \to J/\psi X] \sim 
6 \Gamma[\Lambda_b \to J/\psi X] $. This appears to indicates a breakdown 
of duality unless there is also significant transition of the $\Lambda_b$
 to excited $\Lambda$ which could show up as a $\Sigma \pi$ state.

Phase space enhancements in $\Lambda_b$ decays over $B$ decays 
can be understood in terms of hyperfine interactions \cite{LO,FO,Lipkin}.
In $B\to D^*$ decays there is a phase space disadvantage over 
$B \to D$ transition because of the higher $D^*$ mass but there is a
spin phase space advantage by a factor of three for the $D^*$ in final state
over the $D$ in the final state. So the hyperfine energy is averaged out in 
the $B \to D +D^*$ transition. In $\Lambda_b$ decays isospin conservation 
chooses final states with the lowest hyperfine energy. This added 
hyperfine energy is available for transition and leads to an phase space
 enhancement
in $\Lambda_b$ decays over $B$ decays (A different 
argument \cite{Altarelli} using the scaling of lifetimes as the inverse fifth 
power of hadronic rather than quark masses implicitly gives a larger phase 
space also). Phase space effects were also discussed in a 
different approach in Ref\cite{Chang}. The lesson
from our analysis is that the effect of phase space enhancements  
may be a key factor in
 understanding the lifetime difference 
between $\Lambda_b$ and $B$ hadron.

\vspace{.3in}

\centerline{ {\bf  Acknowledgment}}
We would like to thank M. Neubert for useful comments and discussion.
This work was  supported  in part by grant No.  I-0304-120-.07/93
from The  German-Israeli  Foundation for Scientific  Research and
Development and by the Natural  Sciences and Engineering  Council
of Canada.

\end{document}